\documentclass[prl,twocolumn]{revtex4}
\usepackage{graphicx,amsmath,revsymb,float,bm,times}

\newcommand{\ek}{\epsilon_{\mathbf{k}}}

\newcommand{\Ek}{E_{\mathbf{k}}}

\newcommand{\Omegaq}{\Omega_{\mathbf{q}}}

\newcommand{\uk}{u_{\mathbf{k}}}
\newcommand{\vk}{v_{\mathbf{k}}}

\begin{document}

\title{Density Profiles of Strongly Interacting Trapped Fermi Gases}

\author{Jelena Stajic, Qijin Chen, and K.  Levin}

\affiliation{James Franck Institute and Department of Physics,
  University of Chicago, Chicago, Illinois 60637}

\begin{abstract}
  We study density profiles in trapped fermionic gases, near Feshbach
  resonances, at all $T \leq T_c$ and in the near-BEC and unitary
  regimes.  For the latter, we characterize and quantify the generally
  neglected contribution from noncondensed Cooper pairs. As a
  consequence of these pairs, our profiles are rather well fit to a
  Thomas-Fermi (TF) functional form, and equally well fit to
  experimental data. Our work lends support to the notion that TF fits
  can be used in an experimental context to obtain information about
  the temperature.
\end{abstract}

\maketitle

There is now strong, but not unambiguous, evidence that fermionic
superfluidity has been observed
\cite{Jin4,Ketterle3a,Thomas2a,Grimm3a,Grimm4a,Kinast2} in trapped
atomic gases.  What is particularly exciting about these systems is
that the strength of the pairing interaction can be arbitrarily tuned
(via Feshbach resonance effects) so that superfluidity occurs in a
regime which goes beyond traditional weak coupling BCS theory, towards
Bose-Einstein condensation (BEC) of pre-formed pairs. These
experiments are revealing the nature of superconductivity and
superfluidity in a hitherto unexplored regime.

Experimental proof of superfluidity is not straightforward, except in
the  BEC regime where the particle density profiles acquire a
bi-modal form. This bi-modality is similar to observations in
bosonic atomic systems, which have been systematically studied for the
past decade \cite{RMP}.  In the intermediate
regime between BCS and BEC, the measured density profiles
\cite{Thomas,Grimm2a} in traps do not contain any obvious signatures
of the condensate.  Nevertheless, there is a significant
body of circumstantial evidence in support of superfluidity in this
regime.  This support is based on fast sweep experiments
\cite{Jin4,Ketterle3a}, collective mode measurements
\cite{Thomas2a,Grimm3a,Kinast2}, as well as pairing gap measurements
using radio frequency (RF) techniques \cite{Grimm4a}.

The goals of this paper are to compute the particle density profiles at
all $ T \leq T_c$ (where $T_c$ is the superfluid transition temperature)
in the near-BEC and unitary regimes and to address the implications for
experiment.  For the former our calculation of the contribution from
noncondensed bosons leads to important corrections to estimates of the
condensate fraction based on the conventional Gaussian form. For the
unitary case, these non-condensed pairs 
smooth out the, otherwise abrupt,
transition between the condensate  and the thermal
background of fermionic excitations; this explains why
the measured density profiles
appear to be so featureless \cite{Thomas,Grimm2a}.
%

The unitary or strongly interacting Fermi gas has been the focus of
attention by the community.  Some time ago it was found \cite{Thomas}
that the profiles were reasonably well described by a Thomas-Fermi
(TF) functional form at zero $T$, and in recent work \cite{Kinast}
this procedure has been extended to finite temperatures.
Importantly, there has been no particular theoretical support for these
TF fits. Below $T_c$ all previous theoretical work (which either ignored
non-condensed pair states \cite{JasonHo,Chiofaloa} or used a different
ground state \cite{Strinati4a}) has predicted strong deviations from
this $T$-dependent TF functional form.  It is extremely important, thus,
to have a better understanding of the spatial profiles.
Our work lends theoretical support to the viewpoint \cite{Kinast} that
TF fits (with proper calibration) can be used in an experimental context
to obtain information about the temperature.  In addition to
establishing these TF fits to theory, we compare theory and experiment
directly via one dimensional representations of their respective
profiles.  We demonstrate remarkable agreement in the shapes of the two
profiles.  One can infer from this comparison, experimental support,
albeit indirect, for the presence of a superfluid condensate.


The first generation theories \cite{ourreviewa} have focused on a
BCS-like ground state \cite{Leggett} which is readily generalized to
accommodate a `two channel" variant \cite{JS3} in which there are
Feshbach bosonic (FB) degrees of freedom present as well.  This approach
has met with some initial success in addressing collective mode
experiments \cite{Tosia,Heiselberg} and pairing-gap spectroscopy
\cite{Torma2a}.
The excitations of this standard ground state \cite{JS3,ourreviewa}
consist of two types: noncondensed fermion pairs hybridized with FB, and
Bogoliubov-like fermionic excitations having $\Ek \equiv \sqrt{ (\ek
  -\mu)^2 + \Delta^2 (T) }$, where $\ek$ is the free fermion kinetic
energy.  Note that the ``gap parameter" $\Delta(T)$ is in general
different from the superconducting order parameter,
$\tilde{\Delta}_{sc}$.
Recent RF experiments \cite{Grimm4a} have been analyzed 
\cite{Torma2a} using our formalism to suggest that the
Bogoliubov quasi-particles have an energy gap or pseudogap, which is
present well above $T_c$, and, thus, not directly related to the order
parameter. Additional support for this ``pseudogap" has recently been
reported in another very different
class of experiments \cite{Jin5}.

We summarize the self-consistent equations \cite{JS2a}, in the presence
of a spherical trap, treated at the level of the local density
approximation (LDA) with trap potential $V(r)=\frac{1}{2}m\omega^2 r^2$.
$T_c$ is defined as the highest temperature at which the self-consistent
equations are satisfied precisely at the center.  At a temperature $T$
lower than $T_c$ the superfluid region extends to a finite radius
$R_{sc}$. The particles outside this radius are in a normal state, with
or without a pseudogap.  
%
Our self-consistent equations are given in terms of the Feshbach
coupling constant $g$ and inter-fermion attractive interaction $U$ by a
gap equation
\begin{equation}
1+\left[U+\frac{g^2}{2 \mu-2 V(r)-\nu}\right] \sum_{\bf k} 
\frac{1-2 f(\Ek)}{2 \Ek}=0\,\,,
\label{gap_eq_trap}
\end{equation} 
which is imposed only when $\mu_{pair}(r) = 0$.  The pseudogap
contribution to $\Delta^2(T) = \tilde{\Delta}_{sc}^2(T) +
\Delta_{pg}^2(T)$ is given by
\begin{equation}
\Delta_{pg}^2=\frac{1}{Z} \sum_{\bf q}\, b(\Omega_q -\mu_{pair})\,\,.
\label{eq:1}
\end{equation}
The various residues, $Z$ and $Z_b$, which are of no particular physical
importance here, are described in Ref.~\onlinecite{ourreviewa}.  The
quantity $\Omegaq$ is the pair dispersion, and $\mu_{pair}$ and
$\mu_{boson}$ are the effective chemical potentials of the pairs and FB.

We introduce units $\hbar=1$, fermion mass $m=\frac{1}{2}$, Fermi
momentum $k_F=1$, and Fermi energy $E_F=\hbar \omega (3
N)^{\frac{1}{3}}=1$.
In the figures below, we express length in units of the Thomas-Fermi
radius $R_{TF} =\sqrt {2E_F/(m \omega^2)}=2 (3 N)^{1/3}/k_F $; the
density $n(r)$ and total particle number $N=\int d^3 {\bf r}\, n(r)$ are
normalized by $k_F^3$ and $(k_F R_{TF})^3$, respectively.  The density
of particles at radius $r$ can be written as
\begin{eqnarray}
n(r)&=& 2 n_b^0 + \frac{2}{Z_b} \sum_{\bf q} b(\Omega_q
-\mu_{boson})\nonumber\\ 
&&{}+ 2 \sum_{\bf k}\left [\vk^2 (1-f(\Ek))+\uk^2 f(\Ek)\right] \,\,,
\label{number_eq_trap:above}
\end{eqnarray}
where 
$n_b^0 = g^2 \tilde{\Delta}_{sc}^2/[(\nu -2 \mu +2V(r))U+g^2]^2$ is the
molecular Bose condensate which is only present for $ r \leq R_{sc}$.
The important chemical potential $\mu_{pair} =\mu_{boson}$
\cite{newfootnoteonZg} is identically zero in the superfluid region $r <
R_{sc}$ , and must be solved for self-consistently at larger radii.  Our
calculations proceed by solving numerically the self-consistent
equations.


We now decompose the density profiles into components associated with
the condensate, the noncondensed pairs and the fermions.  This raises a
central issue of this paper. \textit{What is the most suitable
  definition of the condensate ratio in a fermionic superfluid?} Indeed,
in strict BCS theory there are two alternatives \cite{LeggettSB}, one
based on the superfluid density, and the other based on associating the
condensate with the perturbation of the superfluid state relative to the
underlying \textit{free} Fermi gas.  In BCS theory with this second
definition, the zero temperature condensate fraction is of order
$T_c/E_F$, far from the value 100\%, which one obtains from the
superfluid density.  Physically, this second definition reflects the
fraction of the original Fermi liquid states which are modified
substantially by pairing.  We explore both of these here, since they are
expected to enter in different physical contexts.  In the BEC regime
(where fermions are absent) there is no distinction between the two
decompositions, and thus the condensate fraction is uniquely defined.

In the present approach to the BCS-BEC crossover picture it is
relatively straightforward to deduce \cite{ourreviewa} the (local)
superfluid density, $n_s \equiv n_s(r)$.  For the one-channel model we
find \cite{ourreviewa} a simple result for $n_s$, as well as for the
fermionic quasiparticle ($n_{QP}$) and pair contributions ($n_{pair}$)
to the difference $ n - n_s$:
\begin{subequations}
\label{eq:2}
\begin{eqnarray}
n_s&=&\frac{\tilde{\Delta}_{sc}^2}{\Delta^2}n_s^{BCS}(\Delta) \,, \\
n_{pair}&=&  \frac{\Delta_{pg}^2}{\Delta^2}n_s^{BCS}(\Delta) =
n_s^{BCS}(\Delta)-n_s \,,
\end{eqnarray}
\begin{eqnarray}
n_{QP}&=&n-n_s^{BCS}(\Delta) \,,
\end{eqnarray}
\end{subequations}
where $n_s^{BCS}$ is the usual superfluid density for a gas of fermions
within BCS theory, except that here the full excitation gap $\Delta$
appears instead of the order parameter.  In the one-channel case the
order parameter $\tilde{\Delta}_{sc}$ is equal to the Cooper condensate
contribution (which we call $\Delta_{sc}$).  
For the broad Feshbach resonances of $^{40}$K and $^6$Li, these
one-channel results are appropriate for the unitary and BCS regimes of
the two-channel case.  In the BEC regime ($\mu < 0$), the contribution
of the fermionic quasiparticles becomes negligible. Therefore, the
one-channel equations (\ref{eq:2}) will be reduced to
\begin{equation}
n_s = \frac{\tilde{\Delta}_{sc}^2}{\Delta^2} n, \qquad n_{pair} =
\frac{\Delta_{pg}^2}{\Delta^2} n \,. 
\end{equation}
Interestingly, this result is also valid for the two-channel problem in
the BEC regime.


Formally, we can address the second decomposition of the density
profiles by writing the particle density in terms of the single particle
Green's function $G(K)=G_0(K)+G_0(K)\Sigma(K) G(K) $
where $G_0$ represents free fermions, and $K\equiv (i\omega_n,
\mathbf{k})$ is the four-momentum.  The second term is then further
split into the condensed ($\propto \tilde{\Delta}_{sc}^2$) and
noncondensed ($\propto \Delta_{pg}^2$) components. Summing over the Matsubara
frequencies $\omega_n$, and adding the FB contribution
one obtains
\begin{subequations}
\label{eq:3}
\begin{eqnarray}
\tilde{n}_s&=&2 Z \tilde{\Delta}_{sc}^2  \,, \\
\tilde{n}_{pair} &=& 2 \sum_{\bf q\ne 0} b (\Omega_q-\mu_{pair})=2 Z \Delta_{pg}^2 \,,\\
\tilde{n}_{ferm}^{free}&=&2 \sum_{\bf k} f(\ek-\mu+V(r)) \,,
\end{eqnarray}
\end{subequations}
which correspond to condensate density, finite momentum pair/boson
density, and free fermion density, respectively.

\begin{figure}
\includegraphics[width=2.8in,clip]{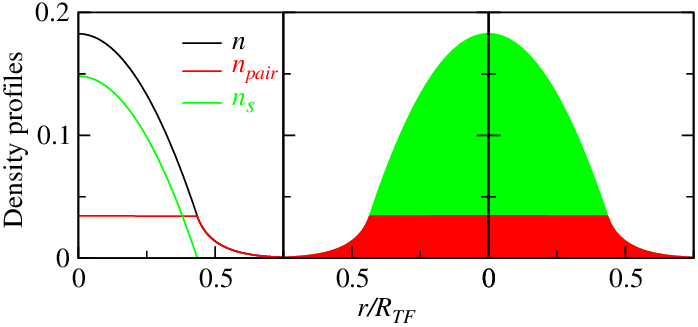}
\caption{(color online) Density profile and its decomposition in the near-BEC limit
  ($1/k_Fa = 3.0$) at $T/T_c = 0.5$.  Shown are the actual plots on the
  left for the total $n(r)$, the noncondensed pairs ${n}_{pair}$, and
  the condensate $n_s$.  The right panel shows how these two (indicated
  by the shaded areas from top to bottom) add together to give $n(r)$.
%
 }
\label{fig:1}
\end{figure}

We characterize the crossover regime by the parameter $1/k_Fa$ where $a$
is the inter-atomic $s$-wave scattering length and $k_F$, the Fermi
momentum.  A typical density profile for the near-BEC limit with $1/k_Fa
= 3.0$ is shown in Fig.~\ref{fig:1} for $T/T_c = 0.5$.  This plot is in
the physically accessible near-BEC regime where the bosons are still
primarily Cooper pairs, as distinguished from Feshbach bosons.  Based on
experiments \cite{Thomas2a,Grimm3a} and related theory
\cite{Tosia,Heiselberg} this is not yet in the deep BEC where the
collective modes are expected to exhibit the characteristics of true
bosons \cite{Stringari}. Thus one expects the ground state wavefunction
to be applicable here.  The left panel corresponds to actual plots of
the condensed (green curve) and noncondensed pair contributions (red)
while the shaded regions on the right represent from top to bottom the
condensed (green) and noncondensed bosons (red), respectively. For this
BEC plot, there are no fermions.  Importantly, the contribution from the
noncondensed bosons is essentially constant in $r$ until $R_{sc}$,
reflecting the fact that they have a gapless excitation spectrum
($\mu_{pair}= 0$).  However, once $\mu_{pair}$ is non-zero at the trap
edge the number of noncondensed bosons (${n}_{pair}$) drops rapidly.

Fig.~\ref{fig:1} is consistent with the experimentally observed profile
shapes \cite{Grimm2a}, but it can be contrasted with other theoretical
results in the literature which predict non-monotonic features
\cite{Strinati4a}.  Because all direct interactions are via fermions,
the character of the BEC profile is different from that of true
interacting bosons.  These differences also appear in the context of
collective mode experiments \cite{Tosia,Heiselberg,Stringari2}.  As for
true bosons \cite{RMP,Griffin4}, the constraint that the bosons are
gapless in the superfluid region is important for determining their
density distribution.
A clear bi-modal feature or ``kink" at $R_{sc}$ is present and provides
evidence for the existence of a condensate.  It should be noted that the
inferred fraction of the condensate may be significantly smaller than
found here -- if a Gaussian form is assumed throughout the trap (as is
the experimental convention) for estimating the contribution from
noncondensed pairs.

\begin{figure}
\includegraphics[width=2.8in,clip]{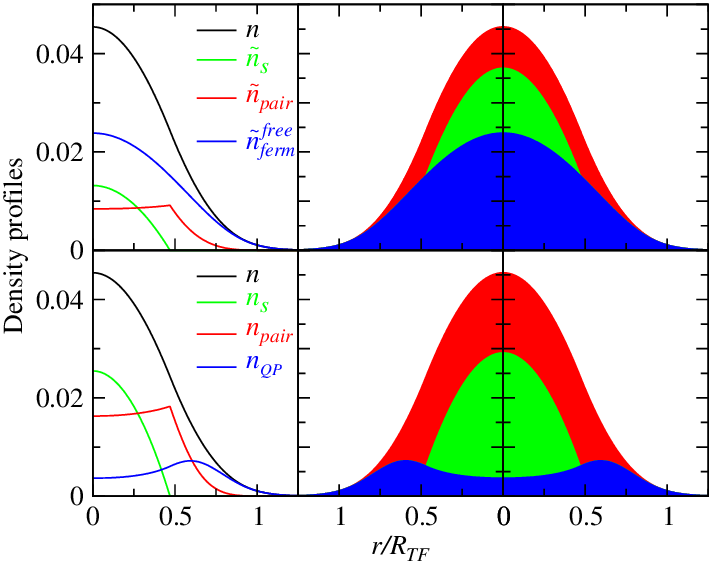}
\caption{(color online) Density profiles and their different decompositions
  for the unitary case ($1/k_Fa=0$) at $T=0.75 T_c$. The upper and
  lower panels correspond to Eqs.~(\ref{eq:3}) and Eqs.~(\ref{eq:2}),
  respectively.  On the left are shown the actual plots for the total,
  the noncondensed pairs, condensate, and fermion density.  Shown on the
  right is how the last three, as indicated by the shaded areas from top
  to bottom of each panel, add up to the local total density $n(r)$.}
\label{fig:2}
\end{figure}

In the upper and lower panels of Fig.~\ref{fig:2} we plot the density
distributions in the unitary limit, $1/k_Fa = 0$, as a function of
radius $r$ at $T/T_c=0.75$, and show their different decompositions
 following Eqs.~(\ref{eq:3}) and Eqs.~(\ref{eq:2}),
respectively.
\textit{The superfluid-based decomposition (lower panels) appears to be more
relevant to thermodynamics and recent RF experiments
\cite{Grimm4a,Torma2a}, since it incorporates the excitation gap of the
fermions}.  The decomposition in the upper panels, based on free
fermions, was studied in detail in Ref.~\onlinecite{Strinati4a}, for a
somewhat different ground state.

The shaded regions on the right of each figure correspond (from top to
bottom) to density profiles of noncondensed pairs, condensate and
fermionic quasiparticles. In both decompositions the noncondensed pairs
play a significant role.  It can be seen from the plots on the left of
Fig.~\ref{fig:2} that the noncondensed pairs show a relatively flat
density distribution for $ r \leq R_{sc}$, just as in the near-BEC limit
of Fig.~\ref{fig:1}.  This behavior similarly arises from the vanishing
of $\mu_{pair}(r)$ in the superfluid region.

One can see that the condensate has
substantially more weight 
for the $n_s$-based decomposition [Eq.~(\ref{eq:2}), Fig.~\ref{fig:2},
lower panels].  Conversely the excited fermions for this case are
significantly less important, since they experience a large gap in their
excitation spectrum.  The contribution of these fermions is concentrated 
at the trap edge.
Indeed, for this case,
because they represent two sides of the same coin,
the fermions and noncondensed bosons
generally appear in different regions of the trap. The latter
are associated with the regions where the excitation gap is
largest (and the fermion states are quasi-bound into 
``bosonic" like states), and the former are found where the
gap is smallest.

\begin{figure}
\includegraphics[width=2.8in,clip]{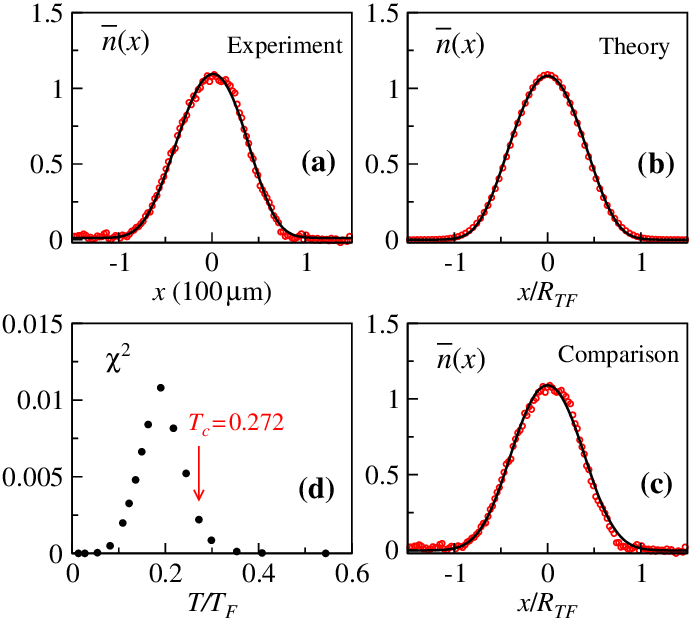}
\caption{(color online) Temperature dependence of (a) experimental
  one-dimensional 
  spatial profiles (circles) and TF fit (line) from
  Ref.~\onlinecite{Kinast}, (b) TF fits (line) to theory at $T \approx
  0.7T_c \approx 0.19T_F$ (circles) and (c) overlay of experimental
  (circles) and theoretical (line) profiles, as well as (d) relative rms
  deviations ($\chi^2$) associated with these fits to theory at
  unitarity. 
The circles in (b) are shown as the line in (c).
  The profiles have been normalized so that $N=\int \bar{n}(x) dx = 1$,
  and we set $R_{TF} = 100~\mu$m in order to overlay the two curves.
  $\chi^2$ reaches a maximum around $T=0.19T_F$.}
\label{fig:4}
\end{figure}

In Fig.~\ref{fig:4} we compare theory (at $T/T_F = 0.19$) and experiment
for the unitary case.  The profiles shown are well within the superfluid
phase ($T_c \approx 0.3T_F$ at unitarity).  This figure presents
Thomas-Fermi fits \cite{Kinast} to the experimental (\ref{fig:4}a) and
theoretical (\ref{fig:4}b) profiles as well as their comparison
(\ref{fig:4}c), for a chosen $R_{TF} = 100~\mu m$, which makes it
possible to overlay the experimental data (circles) and theoretical
curve (line).  Finally Fig.~\ref{fig:4}d indicates the relative $\chi^2$
or root-mean-square (rms) deviations for these TF fits to theory.  This
figure was made in collaboration with the authors of
Ref.~\onlinecite{Kinast}.  This temperature was chosen to exhibit the TF
functional form in the regime where it is most problematic, since
$\chi^2$ has a maximum there. The experimental data were estimated to
correspond to roughly this same temperature, based on the analysis of
Ref.~\onlinecite{Kinast}.  Two of the three dimensions of the
theoretical trap profiles were integrated out to obtain a
one-dimensional representation of the density distribution along the
transverse direction: $\bar {n} (x) \equiv \int dydz \, n(r)$.

This figure is in contrast to earlier theoretical studies which predict
a kink at the condensate edge \cite{JasonHo,Chiofaloa}.  Moreover, in
contrast to the predictions of Ref.~\onlinecite{Strinati4a}, the curves
behave monotonically with both temperature and radius.  Indeed, in the
unitary regime the generalized TF fitting procedure of
Ref.~\onlinecite{Kinast} works surprisingly well for our theory with
spherical traps, and for anisotropic experiments (shown in
Fig.~\ref{fig:4}a).
This is, in part, a consequence of the fact that trap anisotropy effects
become irrelevant for these one dimensional projections.  These
reasonable TF fits apply to essentially all temperatures investigated
experimentally \cite{Kinast}, and all temperatures we have studied,
including in the normal state.
%

To probe the deviations from a TF functional form, and to establish the
role of the condensate, in
Fig.~\ref{fig:4}d, we show the (relative) rms deviation, or $\chi^2$,
from the TF fits as a function of $T$.  Except at the very lowest
temperatures (where the $\chi^2$ is small and the condensate occupies
the entire region of the trap), the condensate causes a small,
but potentially measurable variation from the TF form.
 $\chi^2$ increases rapidly below $T_c$ and
reaches a maximum around $0.7T_c$.  Here the profile involves two
different functional forms: the condensate and the thermal distribution
functions. More precisely, the systematic behavior of $\chi^2$ indicates
when $T$ has safely surpassed $T_c$.

The reason the Thomas-Fermi fits work well below $T_c$ has to do with
the presence of noncondensed pairs.  If these bosonic-like states are
ignored, (by omitting the top-most shaded regions) in the right panels
of Fig.~\ref{fig:2}, a clear bi-modal distribution emerges, as has been
predicted in the literature\cite{Chiofaloa,JasonHo}. In this case, TF
fits would not work well. A residue of this previously discussed
bi-modality must be present to some extent in the present plots, but
this can only be seen in the derivatives of our profiles.  We find a
small kink in the first order derivative and a sharp jump in the second
order derivative of $n(r)$, reflecting the condensate edge.  These
features may be more difficult to identify with current experimental
techniques. Alternatively, one may infer the presence of a condensate
through $\chi^2$ plots from TF fits.
These experiments will require substantial averaging over a large number
of profiles, however.

In summary, in this paper we have found that, in the near BEC regime,
the component profile for the noncondensed bosons, is different from the
Gaussian form generally assumed, although the general shape is
consistent with experiment. This result which follows because these
noncondensed states are in equilibrium with a condensate, should have
implications for measurements of the condensate fraction. Near
unitarity, we have found that, our calculated density profiles are
consistent with experiment and provide strong support for using Thomas
Fermi fits to the profiles. One can infer that the condensate is seen,
not in the profile shapes, but, presumably, in their temperature
calibration.


We acknowledge very substantial help from J. E. Thomas, J. Kinast and A.
Turlapov with the Thomas-Fermi analysis in this paper, and for sharing
their experimental data. Useful discussions with M. Greiner, D.S. Jin
and C. Chin are gratefully acknowledged.  This work was supported by
NSF-MRSEC Grant No.~DMR-0213745.


\begin{thebibliography}{26}
\bibitem[{new()}]{newfootnoteonZg}
\bibinfo{note}{Outside the superfluid region, there are small but unimportant
  differences between $\mu_{pair}$ and $\mu_{boson}$.}
\end{thebibliography}

\end{document}